\documentclass[
aps, 
pre,
amsmath,amssymb,
reprint,%
]{revtex4-1}
\usepackage{graphicx}
\usepackage{dcolumn}
\usepackage{bm}

\usepackage{amsmath}	
\begin{document}

\newcommand{\diff}[2]{\frac{d#1}{d#2}}
\newcommand{\pdiff}[2]{\frac{\partial #1}{\partial #2}}
\newcommand{\fdiff}[2]{\frac{\delta #1}{\delta #2}}
\newcommand{\bx}{\bm{x}}
\newcommand{\bq}{\bm{q}}
\newcommand{\br}{\bm{r}}
\newcommand{\bu}{\bm{u}}
\newcommand{\by}{\bm{y}}
\newcommand{\bY}{\bm{Y}}
\newcommand{\bF}{\bm{F}}
\newcommand{\new}{\nonumber\\}
\newcommand{\abs}[1]{\left|#1\right|}
\newcommand{\tr}{{\rm Tr}}
\newcommand{\ave}[1]{\left\langle #1 \right\rangle}
\newcommand{\im}{{\rm Im}}
\newcommand{\re}{{\rm Re}}
\newcommand{\dr}{\delta r}
\newcommand{\tz}{\tilde{z}}
\newcommand{\txi}{\tilde{\xi}}

\preprint{AIP/123-QED} \title{Solvable model of noisy coupled
oscillators with fully random interactions}

\author{Harukuni Ikeda}
 \email{harukuni.ikeda@yukawa.kyoto-u.ac.jp}
\affiliation{Yukawa Institute for Theoretical Physics, Kyoto University,
Kyoto 606-8502, Japan}

\date{\today}

\begin{abstract}
We introduce a solvable spherical model of coupled oscillators with
fully random interactions and distributed natural frequencies. Using the
dynamical mean-field theory, we derive self-consistent equations for the
stationary response and correlation functions. We show that the
finite-temperature spin-glass transition is suppressed for any
nonmonodisperse natural-frequency distribution, including continuous,
discrete, and asymmetric distributions, because the resulting
low-frequency singularity of the correlation function is incompatible
with the spherical constraint. At zero temperature, however, a
spin-glass phase persists for arbitrary frequency dispersion. This
residual zero-temperature glassiness is likely a special feature of the
spherical dynamics and would be destroyed by local nonlinearities. The
model thus provides a solvable oscillator framework for studying how
nonequilibrium perturbations suppress finite-temperature glassy
freezing.
\end{abstract}

\maketitle

\section{Introduction}

Synchronization is a ubiquitous collective phenomenon observed in a wide
variety of nonequilibrium systems, ranging from biological and chemical
oscillators to engineered dynamical networks. A standard theoretical
framework for studying synchronization is the Kuramoto model, which
describes interacting phase oscillators with distributed natural
frequencies~\cite{Kuramoto1984}. Owing to its simplicity and analytical
accessibility, the Kuramoto model and its variants have played a central
role in the study of collective dynamics in many-body
systems~\cite{Strogatz2000,Acebrn2005,Rodrigues2016}.

An important extension of the Kuramoto model is to introduce disorder
into the
interactions~\cite{daido1992,kloumann2014,iatsenko2014glassy,daido2018,ottino2018,kimoto2019,pazo2023,pruser2024,pikovsky2024,leon2025dynamics,juhasz2025}.
When the couplings contain both positive and negative components,
frustration arises, and the system may exhibit glassy collective
behavior in addition to synchronization. Randomly coupled oscillator
systems have therefore attracted considerable interest as nonequilibrium
analogues of disordered mean-field systems in equilibrium statistical
mechanics. Exact analytical results, however, are available only for
special choices of the interaction matrix, such as low-rank
couplings~\cite{bonilla1993glassy,kloumann2014,ottino2018,iatsenko2014glassy,pazo2023},
whereas models with fully random interactions have been studied mainly
by numerical or perturbative
approaches~\cite{daido1992,pruser2024,leon2025dynamics}.  Recent works
suggest that a robust glass phase may be absent in
the thermodynamic limit~\cite{leon2025dynamics}, but the theoretical
basis for this conclusion remains incomplete because these models are
not analytically solvable.

A useful perspective on this problem comes from earlier studies of
nonequilibrium spin-glass
models~\cite{sherrington1975,kosterlitz1976,crisanti1987,
cugliandolo1993,cugliandolo1995, giulia2025prl,lorenzana2025pre}. In
particular, Crisanti and Sompolinsky showed that, in spin-glass models
with asymmetric interactions, perturbations away from the equilibrium
limit destroy the finite-temperature spin-glass
phase~\cite{crisanti1987}. This raises the question of whether a related
mechanism may also operate in disordered oscillator systems, where
nonequilibrium effects arise not from asymmetric couplings but from
distributed natural frequencies.

In this paper, we introduce a solvable mean-field model of randomly
coupled oscillators inspired by the previous works of the spherical
spin-glass
model~\cite{berlin1952,kosterlitz1976,crisanti1987,cugliandolo1993,cugliandolo1995,castellani2005}
and its extension for complex variables~\cite{antenucci2015}.  In the
original Kuramoto model, each oscillator is represented by a phase
variable $\theta_i$, or equivalently by a complex amplitude
$z_i=e^{i\theta_i}$ with fixed unit modulus, $\abs{z_i}^2=1$. In our
model, these local constraints are relaxed and replaced by a global
spherical constraint, $\sum_{i=1}^N \abs{z_i}^2=N$. This replacement
renders the model analytically tractable and allows us to derive closed
self-consistent equations for the stationary response and correlation
functions.

Without any amplitude constraint, the model reduces to a linear Gaussian
theory. Although this theory is analytically solvable, an instability
leads to unbounded growth rather than to a stable ordered or glassy
phase. The local unit-modulus constraints of the Kuramoto model provide
the saturation required beyond such an instability, but make the fully
random problem analytically difficult. The global spherical constraint
provides an intermediate model between these two limits: it stabilizes
the total amplitude while preserving analytical tractability.  Such
spherical constructions have long been used in equilibrium statistical
mechanics to construct solvable models of ferromagnetic and spin-glass
transitions~\cite{berlin1952,kosterlitz1976,crisanti1987,cugliandolo1993,cugliandolo1995,castellani2005,antenucci2015}. Here,
it allows us to isolate the effect of natural-frequency dispersion on
the glass instability.  The resulting model should therefore be viewed
as a solvable reference model rather than as a quantitatively faithful
approximation to the locally constrained Kuramoto model.

As a first step, we show that the spherical formulation reproduces the
basic synchronization transition in the ferromagnetic case. Although the
spherical model is simpler than the original phase model, it still
captures the standard synchronization transition in this benchmark
setting.

We then turn to fully random Gaussian couplings. Using the dynamical
mean-field
theory~\cite{crisanti1987,cugliandolo1993,castellani2005,giulia2025prl},
we derive closed self-consistent equations for the stationary response
and correlation functions, which allow us to examine the possibility of
a spin-glass transition in the presence of distributed natural
frequencies. We find that a finite-temperature spin-glass transition
occurs only in the singular limit where all natural frequencies are
identical. In this limit, the model reduces to the spherical
Sherrington-Kirkpatrick model and reproduces its
transition~\cite{sherrington1975,kosterlitz1976}. Once the frequency
distribution has a finite width or multiple delta peaks, however, the
finite-temperature transition is suppressed, because the resulting
low-frequency singularity of the correlation function is incompatible
with the spherical constraint. At zero temperature, by contrast, a
spin-glass phase persists even for finite frequency dispersion. This
residual zero-temperature glassiness is likely a special feature of the
spherical dynamics and would be removed by local
nonlinearities~\cite{crisanti1987}.

This paper is organized as follows. In Sec.~\ref{120945_24Mar26}, we
review the Kuramoto model and introduce its spherical version. In
Sec.~\ref{120957_24Mar26}, we discuss the model with ferromagnetic
interaction as a benchmark case. In Sec.~\ref{125859_24Mar26}, we
investigate the stationary regime of the model with fully random
interactions by means of the dynamical mean-field theory.
Section~\ref{125933_24Mar26} is devoted to summary and discussions.

\section{Model}
\label{120945_24Mar26}

In this section, we introduce the model studied in this work.
We begin with the standard Kuramoto model and then formulate
its spherical counterpart, which replaces the local unit-modulus
constraints by a global spherical constraint. This modification
preserves the basic oscillator structure while rendering the model
analytically tractable.

\subsection{Kuramoto model}

The Kuramoto model is defined by the equation of motion~\cite{Kuramoto1984,Strogatz2000,Acebrn2005}
\begin{align}
 \dot{\theta}_i = \Omega_i + \sum_{j=1}^N J_{ij}\sin(\theta_j-\theta_i),
\end{align}
where $\Omega_i$ denotes the natural frequency and $J_{ij}$ is a
symmetric interaction matrix. For later convenience, we rewrite the
dynamics in terms of the complex amplitude
\begin{align}
 z_i = e^{i\theta_i}.
\end{align}
In this representation, the equation of motion can be written
as~\cite{yamaguchi1984,matthews1990}
\begin{align}
\dot{z}_i = -\mu_i z_i + i\Omega_i z_i + \sum_{j=1}^N J_{ij}z_j,
\label{170649_6Jan26}
\end{align}
where $\mu_i$ is a Lagrange multiplier enforcing the constraint
$\abs{z_i}^2=1$. The difficulty is that $\mu_i$ depends implicitly on
the instantaneous state, which makes the dynamics nonlinear and
precludes a simple analytical treatment.

\subsection{Spherical model}

To obtain an analytically solvable model, we relax the local constraints
$\abs{z_i}^2=1$ and instead impose the global spherical constraint
\begin{align}
 \sum_{i=1}^N \abs{z_i}^2 = N.
\label{213913_6Jan26}
\end{align}
The equation of motion then becomes
\begin{align}
\dot{z}_i = -\mu z_i + i\Omega_i z_i + \sum_{j=1}^N J_{ij}z_j + \xi_i,
\label{091554_28Feb26}
\end{align}
where $\mu$ is the Lagrange multiplier enforcing
Eq.~(\ref{213913_6Jan26}). We also include a complex Gaussian white
noise with zero mean and covariance~\cite{antenucci2015}
\begin{align}
\ave{\xi_i(t)\xi_j^*(t')} = 2T\delta_{ij}\delta(t-t'),
\qquad
\ave{\xi_i(t)\xi_j(t')} = 0.
\end{align}

This spherical formulation retains the competition among coupling,
frequency disorder, and noise, while greatly simplifying the analysis.
As we show below, it reproduces the standard synchronization transition
in the ferromagnetic case and remains analytically tractable even in the
presence of fully random interactions.

The Lagrange multiplier $\mu$ is determined from the spherical
constraint~\cite{cugliandolo1993,cugliandolo1995}. Taking the time
derivative of Eq.~(\ref{213913_6Jan26}), we obtain
\begin{align}
\diff{}{t}\sum_{i=1}^N \abs{z_i}^2
= -2\mu N + 2\sum_{ij}J_{ij}z_iz_j^* + 2TN
= 0,
\end{align}
which yields
\begin{align}
\mu = \frac{1}{N}\sum_{ij}J_{ij}z_iz_j^*+T.
\label{233612_31Mar26}
\end{align}
This expression shows in particular that $\mu$ is real.

\section{Benchmark case: ferromagnetic interaction}
\label{120957_24Mar26}

Before turning to fully random interactions, we first consider the
ferromagnetic case as a benchmark. In this case, the spherical model
reproduces the standard synchronization transition and thus provides a
useful reference point for the disordered model studies below.

\subsection{Settings}
We consider the uniform interaction matrix
\begin{align}
J_{ij} = \frac{K}{N}.
\end{align}
The equation of motion then becomes 
\begin{align}
\dot{z}_i(t) = -\mu z_i(t) + i\Omega_i z_i(t) + KZ(t)+\xi_i,\label{111936_8Jan26}
\end{align}
where 
\begin{align}
Z(t) = \frac{1}{N}\sum_{i=1}^N z_i(t)\label{123901_7Jan26}
\end{align}
is the complex order parameter.  The state with ${Z=0}$ corresponds to
an incoherent state, whereas ${\abs{Z}>0}$ signals synchronization.

\subsection{Steady state}
We first move to a uniformly rotating frame in which the mean natural
frequency vanishes. We then analyze the steady state, in which $\mu$ and
$Z$ are time independent. Since Eq.~(\ref{111936_8Jan26}) is linear, it
can be solved straightforwardly by Fourier transformation:
\begin{align}
z_i(\omega) = 2\pi\delta(\omega) \frac{KZ}{\mu-i\Omega_i}
 + \frac{\xi_i(\omega)}{\mu+i(\omega-\Omega_i)}.
\end{align}
Here and below, Fourier-transformed quantities are written as functions
of $\omega$, with the convention
\begin{align}
f(\omega) = \int_{-\infty}^\infty dt  e^{-i\omega t}f(t).
\end{align}
Assuming that $\xi_i$ and $\Omega_i$ are uncorrelated,
and averaging over $i=1,\cdots, N$, we obtain
\begin{align}
 Z = \frac{KZ}{N}\sum_{i=1}^N \frac{1}{\mu-i\Omega_i}
 = KZ \int d\Omega g(\Omega)\frac{1}{\mu-i\Omega},\label{140621_25Mar26}
\end{align}
where 
\begin{align}
 g(\Omega) = \frac{1}{N}\sum_{i=1}^N \delta(\Omega-\Omega_i)
\end{align}
is the distribution of natural frequencies.
The Lagrange multiplier~(\ref{233612_31Mar26}) reduces to 
\begin{align}
\mu = K\abs{Z}^2 + T.\label{144937_25Mar26}
\end{align}
The order parameter is determined self-consistently from
Eqs.~(\ref{140621_25Mar26}) and (\ref{144937_25Mar26}).

\begin{figure}[t]
\begin{center}
 \includegraphics[width=8cm]{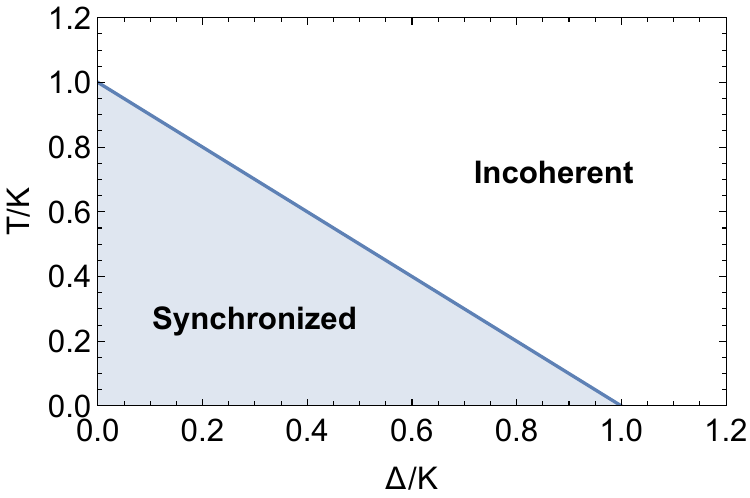} \caption{ Phase diagram of the
spherical model with ferromagnetic interactions. The filled and unfilled
regions denote the synchronized and incoherent phases, respectively.}
\label{110035_1Apr26}
\end{center}
\end{figure}

Unless otherwise stated, we assume in the following that the natural
frequencies are drawn from the Cauchy distribution
\begin{align}
g(\Omega) =
\frac{\Delta}{\pi}\frac{1}{\Omega^2+\Delta^2}.\label{114928_26Mar26}
\end{align}
In this case, Eq.~(\ref{140621_25Mar26}) becomes 
\begin{align}
 Z = \frac{KZ}{\mu+\Delta}.\label{234309_31Mar26}
\end{align}
This result also follows directly from the analytic structure of the
Cauchy distribution. Because $g(\Omega)$ has a simple pole at
$\Omega=i\Delta$ in the upper half-plane, the integral in
Eq.~(\ref{140621_25Mar26}) is given by the residue at that pole, namely
by evaluating the non-singular part of the integrand at
$\Omega=i\Delta$~\cite{ott2008}. Combining Eqs.~(\ref{144937_25Mar26})
and (\ref{234309_31Mar26}), we obtain
\begin{align}
\abs{Z}=
\begin{cases}
  0 & T> T_c,\\ 
 \sqrt{\frac{K-T-\Delta}{K}} & T\leq Tc,
\end{cases} 
\end{align}
with the transition temperature 
\begin{align}
 T_c = K-\Delta.
\end{align}
The resulting phase diagram is shown in Fig.~\ref{110035_1Apr26}. The
spherical model reproduces the standard mean-field synchronization
transition in the presence of both frequency dispersion and thermal
noise, in qualitative agreement with previous results for noisy Kuramoto
models~\cite{Kuramoto1984,Sakaguchi1988,Strogatz1991,Acebrn2005}.  The
same result can also be obtained by directly analyzing the relaxation
dynamics of the ferromagnetic model, see Appendix.~\ref{180637_3Apr26}.

\section{Random interactions}
\label{125859_24Mar26}

\subsection{Settings}

We now consider random interactions. The coupling matrix $J_{ij}$ is
symmetric, $J_{ij}=J_{ji}$, and is drawn from a Gaussian distribution
with zero mean and covariance
\begin{align}
\ave{J_{ij}J_{lk}} = \frac{J^2}{N}\left(\delta_{il}\delta_{jk}+\delta_{ik}\delta_{jl}\right).
\label{112022_28Feb26}
\end{align}
Since the model is fully connected, its dynamics can be treated by a
standard cavity construction, or equivalently by a dynamical mean-field
theory for the single-site response and correlation functions
\cite{giulia2025prl,lorenzana2025pre,blumenthal2025building}.

The present model is closely related to the nonreciprocally coupled
spherical spin-glass models studied in
Refs.~\cite{giulia2025prl,lorenzana2025pre}. Indeed, writing
$z_i=s_i^2+i s_i^1$, the natural frequency $\Omega_i$ appears as a
site-dependent antisymmetric coupling between the two real
components. The main difference lies in the constraint structure:
Refs.~\cite{giulia2025prl,lorenzana2025pre} impose separate spherical
constraints on the two components, whereas the present model imposes a
single constraint on their total norm. The stationary dynamical
mean-field equations are closely related when the two Lagrange
multipliers of the former model
coincide. Reference~\cite{lorenzana2025pre} also found that a continuous
distribution of antisymmetric couplings suppresses the
finite-temperature transition through a nonintegrable singularity of the
correlation function. However, the predictions differ for discrete
distributions with multiple peaks. Ref.~\cite{lorenzana2025pre} finds a
finite-temperature transition, whereas in the present model it occurs
only in the monodisperse limit. The latter reduces to the equilibrium
spherical Sherrington--Kirkpatrick model in a rotating frame. This
difference demonstrates that the two constraint structures are not
generally equivalent.

\subsection{Cavity equations}

We now derive the corresponding effective single-site process. The key
step is to add an extra variable $z_0(t)$ to the original $N$-site
system and examine its coupling to the other degrees of freedom. In the
thermodynamic limit $N\to\infty$, the enlarged $(N+1)$-site system is
statistically equivalent to the original one, which closes the cavity
construction.

For $i=1,\dots,N$, the equation of motion becomes
\begin{align}
\dot{z}_i(t)
=
-\mu z_i(t)+i\Omega_i z_i(t)
+\sum_{j=1}^N J_{ij}z_j(t)
+\xi_i(t)+\delta\xi_i(t),
\end{align}
where
\begin{align}
\delta\xi_i(t)=J_{i0}z_0(t)
\end{align}
represents the perturbation induced by the additional site $z_0$. To
linear order in this perturbation, one obtains
\begin{align}
z_i(t)
=
z_{i\backslash 0}(t)
+\sum_{j=1}^N\int_{-\infty}^t dt'\,
R_{ij}(t,t')\,\delta\xi_j(t'),
\end{align}
up to corrections of order $1/N$. Here $z_{i\backslash 0}(t)$ denotes
the dynamical variable in the cavity system without site $0$, and
\begin{align}
R_{ij}(t,t')=\fdiff{z_i(t)}{\xi_j(t')}
\end{align}
is the response function. 

The equation of motion for the added site
reads
\begin{align}
\dot{z}_0(t)
=
-\mu z_0(t)
+i\Omega_0 z_0(t)
+\xi_0(t)
+\sum_{i=1}^N J_{0i}z_i(t),
\label{112132_28Feb26}
\end{align}
where $J_{0i}$ obeys the same statistics as the original couplings:
$J_{0i}=J_{i0}$, $\ave{J_{0i}}=0$, and
$\ave{J_{0i}J_{0j}}=\delta_{ij}J^2/N$. Substituting the linear-response
expression for $z_i(t)$ into the last term of
Eq.~(\ref{112132_28Feb26}), we obtain~\cite{blumenthal2025building}
\begin{align}
\sum_{i=1}^N J_{0i}z_i(t)
&=
\sum_{i=1}^N J_{0i}z_{i\backslash 0}(t)\new 
&+
\int_{-\infty}^t dt' \sum_{ij}
J_{0i}R_{ij}(t,t')J_{j0}z_0(t')
\notag\\
&\approx
\sum_{i=1}^N J_{0i}z_{i\backslash 0}(t)
+
J^2\int_{-\infty}^t dt'\,R(t,t')z_0(t'),
\label{112122_28Feb26}
\end{align}
where
\begin{align}
R(t,t')=\frac{1}{N}\sum_{i=1}^N R_{ii}(t,t').
\end{align}
To justify the second term in Eq.~(\ref{112122_28Feb26}), note that
$R_{ij}(t,t')$ is evaluated in the cavity system without site $0$ and is
therefore independent of $J_{0i}$ and $J_{j0}$ to leading order in
$1/N$. In the thermodynamic limit, this quantity is self-averaging.
 One may
then average over the couplings to obtain
\begin{align}
\ave{\sum_{ij}J_{0i}R_{ij}(t,t')J_{j0}}
=
\frac{J^2}{N}\sum_i R_{ii}(t,t')
=
J^2R(t,t'),
\end{align}
which justifies the replacement in
Eq.~(\ref{112122_28Feb26})~\cite{blumenthal2025building}.  We thus
arrive at the effective single-site process
\begin{align}
\dot{z}_0(t)
&=
-\mu z_0(t)
+i\Omega_0 z_0(t)
+J^2\int_{-\infty}^t dt'\,R(t,t')z_0(t')
\notag\\
&\quad
+\xi_0(t)+\eta_0(t),
\label{182348_26Mar26}
\end{align}
where
\begin{align}
\eta_0(t)=\sum_{i=1}^N J_{0i}z_{i\backslash 0}(t)
\end{align}
is a sum of many weak random contributions and therefore becomes an
effective Gaussian noise with covariance
\begin{align}
\ave{\eta_0(t)\eta_0^*(t')}
&=J^2C(t,t'),
\notag\\
C(t,t')
&=
\frac{1}{N}\sum_{i=1}^N
\ave{z_{i\backslash 0}(t)z_{i\backslash 0}^*(t')}.
\end{align}
The structure of the effective stochastic process derived above is
essentially the same as that obtained previously for the spherical SK
model and related mean-field disordered
systems~\cite{crisanti1987,castellani2005,blumenthal2025building,giulia2025prl},
although in the present case it is extended to complex variables with
distributed natural frequencies.

\subsection{Stationary regime}
In the spherical SK model, the low-temperature dynamics exhibits aging
and is not fully time-translationally invariant. Nevertheless, a
well-defined stationary limit of the response and correlation functions
exists. We restrict the following analysis to the stationary regime and
do not consider the aging regime. It corresponds to the limit in which
the waiting time tends to infinity at a fixed time difference $\tau =
t-t'$~\cite{cugliandolo1995}:
\begin{align}
R_{\rm st}(\tau)
&=
\lim_{t_{\rm w}\to\infty}
R(t_{\rm w}+\tau,t_{\rm w}),
\\
C_{\rm st}(\tau)
&=
\lim_{t_{\rm w}\to\infty}
C(t_{\rm w}+\tau,t_{\rm w}).
\end{align}
These functions depend only on the time difference and can therefore be analyzed in
Fourier space. The aging regime, in which $\tau=O(t_{\rm w})$, is not
described by the following equations. For notational simplicity, we
omit the subscript ``st'' below.

For a fixed value of $\Omega_0$, we denote the corresponding solution by
$z_0(t|\Omega_0)$. Upon averaging over $\Omega_0$ with weight
$g(\Omega_0)$, the added site becomes statistically equivalent to a
typical site in the original system in the thermodynamic limit.  Fourier
transforming the time-translation-invariant equations, we obtain
\begin{align}
z_0(\omega|\Omega_0)
=
R_0(\omega|\Omega_0)\left[\xi_0(\omega)+\eta_0(\omega)\right],
\label{205027_26Mar26}
\end{align}
where we have defined the response for $z_0(\omega|\Omega_0)$ as follows:
\begin{align}
R_0(\omega|\Omega_0)
= \pdiff{z_0(\omega|\Omega_0)}{\xi_0(\omega)}
= \frac{1}{i\omega-i\Omega_0+\mu-J^2R(\omega)}.
\end{align}
Averaging over $\Omega_0$, we then obtain the self-consistency equation
\begin{align}
R(\omega)
=
\int d\Omega_0\, g(\Omega_0)R_0(\omega|\Omega_0).
\label{110536_2Mar26}
\end{align}
This equation determines the response function $R(\omega)$. Similarly,
taking the disorder and noise average of the squared modulus of
Eq.~(\ref{205027_26Mar26}), we obtain
\begin{align}
C_0(\omega|\Omega_0)
=
\abs{R_0(\omega|\Omega_0)}^2
\left[2T+J^2C(\omega)\right],
\end{align}
where $C_0(\omega|\Omega_0)$ is the Fourier transform of the single-site
correlation function for fixed $\Omega_0$. Averaging over $\Omega_0$
yields
\begin{align}
C(\omega)
=
\left(2T+J^2C(\omega)\right)
\int d\Omega_0\, g(\Omega_0)\abs{R_0(\omega|\Omega_0)}^2.
\label{215733_2Mar26}
\end{align}
The integral in Eq.~(\ref{215733_2Mar26}) can be evaluated as
\begin{align}
&\int d\Omega_0\, g(\Omega_0)\abs{R_0(\omega|\Omega_0)}^2\new 
&=
\int d\Omega_0\, g(\Omega_0)
\frac{1}{s(\omega)+s^*(\omega)}
\left(
\frac{1}{s(\omega)-i\Omega_0}
+
\frac{1}{s^*(\omega)+i\Omega_0}
\right)
\new 
&=
\frac{R(\omega)+R^*(\omega)}{s(\omega)+s^*(\omega)}
\new 
&=
\frac{\alpha(\omega)}{\mu-J^2\alpha(\omega)},
\end{align}
where $\alpha(\omega)=\mathrm{Re}\,R(\omega)$ and
\begin{align}
s(\omega)=i\omega+\mu-J^2R(\omega).
\end{align}
Substituting this result into Eq.~(\ref{215733_2Mar26}), we find
\begin{align}
C(\omega)
=
\frac{2T\alpha(\omega)}{\mu-2J^2\alpha(\omega)}.
\label{112733_28Mar26}
\end{align}
Finally, the Lagrange multiplier $\mu$ is fixed by the spherical
constraint,
\begin{align}
1
=
\frac{1}{2\pi}\int d\omega\, C(\omega)
=
\frac{T}{\pi}\int d\omega\,
\frac{\alpha(\omega)}{\mu-2J^2\alpha(\omega)}.
\label{105341_3Mar26}
\end{align}

\subsection{Edwards--Anderson parameter and transition point}

We now examine the possible emergence of a nonzero plateau in the
stationary regime defined in the previous subsection. The
Edwards--Anderson parameter is defined by taking the waiting time to
infinity before the time difference:
\begin{align}
C_{\infty}
\equiv
\lim_{\tau\to\infty}C(\tau)
=\lim_{\tau\to\infty}
\lim_{t_{\rm w}\to\infty}
C(t_{\rm w}+\tau,t_{\rm w})
>0.
\label{eq:qEA_definition}
\end{align}
Equivalently, the Fourier transform of the stationary correlation
function contains a delta-function contribution at zero frequency,
\begin{align}
C(\omega)
=
\delta C(\omega)
+
2\pi C_{\infty}\delta(\omega),
\label{eq:C_decomposition_qEA}
\end{align}
where $\delta C(\omega)$ denotes the regular part. Substituting this
decomposition into Eq.~(\ref{215733_2Mar26}) and collecting the terms
proportional to $C_\infty$, we obtain
\begin{align}
1
&=
J^2 \int d\Omega_0\, g(\Omega_0)\abs{R_0(0|\Omega_0)}^2
\notag\\
&=
\frac{J^2\alpha(0)}{\mu-J^2\alpha(0)},
\label{151139_7Mar26}
\end{align}
which implies
\begin{align}
\mu = 2J^2\alpha(0).
\label{021141_3Mar26}
\end{align}
Substituting this relation into the spherical
constraint~(\ref{105341_3Mar26}), we obtain the transition temperature
\begin{align}
T_c
=
2\pi J^2
\left(
\int d\omega\,
\frac{\alpha(\omega)}{\alpha(0)-\alpha(\omega)}
\right)^{-1}.
\label{145324_5Mar26}
\end{align}

For $T<T_c$, the delta-function contribution at $\omega=0$ must be
treated explicitly in the spherical constraint:
\begin{align}
1
=
C_\infty
+
\frac{T}{\pi}\int d\omega\,
\frac{\alpha(\omega)}{\mu-2J^2\alpha(\omega)}.
\end{align}
Using Eq.~(\ref{021141_3Mar26}) and (\ref{145324_5Mar26}), we then find
\begin{align}
C_\infty = 1-\frac{T}{T_c}.
\end{align}

The quantity $C_{\infty}$ characterizes the plateau reached in the
stationary regime. The subsequent aging decay on time scales
$\tau=O(t_{\rm w})$ is not described by the present Fourier-space
analysis~\cite{cugliandolo1995}.

\subsection{Cauchy distribution}

We now specialize to the Cauchy distribution,
\begin{align}
g(\Omega)=\frac{\Delta}{\pi}\frac{1}{\Omega^2+\Delta^2},
\label{114928_26Mar26}
\end{align}
for which the self-consistency equation for the response function can be
solved analytically. 
Starting from Eq.~(\ref{110536_2Mar26}), and proceeding as in the
ferromagnetic case, the integral over $\Omega$ can be evaluated by
contour integration for the Cauchy distribution, yielding
\begin{align}
R(\omega)
=
\frac{1}{i\omega+\Delta+\mu-J^2R(\omega)},
\end{align}
leading to 
\begin{align}
R(\omega)
=
\frac{i\omega+\Delta+\mu-\sqrt{(i\omega+\Delta+\mu)^2-4J^2}}{2J^2}.
\end{align}
Its real part, $\alpha(\omega)=\mathrm{Re}\,R(\omega)$, is given by
\begin{align}
\alpha(\omega)
=
\frac{1}{2J^2}
\left[
\Delta+\mu
-\sqrt{\frac{\sqrt{a(\omega)^2+b(\omega)^2}+a(\omega)}{2}}
\right],
\label{105639_3Mar26}
\end{align}
where
\begin{align}
a(\omega)=(\Delta+\mu)^2-\omega^2-4J^2,
\qquad
b(\omega)=2\omega(\Delta+\mu).
\end{align}
For finite $\Delta$, the small-$\omega$ behavior is
\begin{align}
\alpha(\omega)
=
\alpha(0)
-\frac{\omega^2}{\left[(\Delta+\mu)^2-4J^2\right]^{3/2}}
+O(\omega^3).
\label{112159_7Mar26}
\end{align}
The transition condition, Eq.~(\ref{021141_3Mar26}),
determines the critical value of the Lagrange multiplier as
\begin{align}
\mu_c=-\Delta+\sqrt{\Delta^2+4J^2}.
\label{112203_7Mar26}
\end{align}

We can then examine whether a finite-temperature transition exists. For
$\Delta=0$, the equations reduce to those of the spherical
Sherrington-Kirkpatrick model~\cite{castellani2005}, and one recovers
the standard result
\begin{align}
T_c=J.
\end{align}
For any finite $\Delta$, however, the situation changes qualitatively.
At the putative transition point, Eqs.~(\ref{112159_7Mar26}) and
(\ref{112203_7Mar26}) imply
\begin{align}
\frac{\alpha(\omega)}{\alpha(0)-\alpha(\omega)}
\sim
\frac{\alpha(0)\Delta^3}{\omega^2}
\qquad
(\omega\to 0).
\end{align}
The integral in Eq.~(\ref{145324_5Mar26}) therefore diverges in the
infrared, so that the transition temperature is driven to zero:
\begin{align}
T_c=0
\qquad
(\Delta>0).
\end{align}
Thus, within the present spherical dynamics, any finite width of the
natural-frequency distribution destroys the finite-temperature
spin-glass transition.

\begin{figure}[t]
\begin{center}
 \includegraphics[width=9cm]{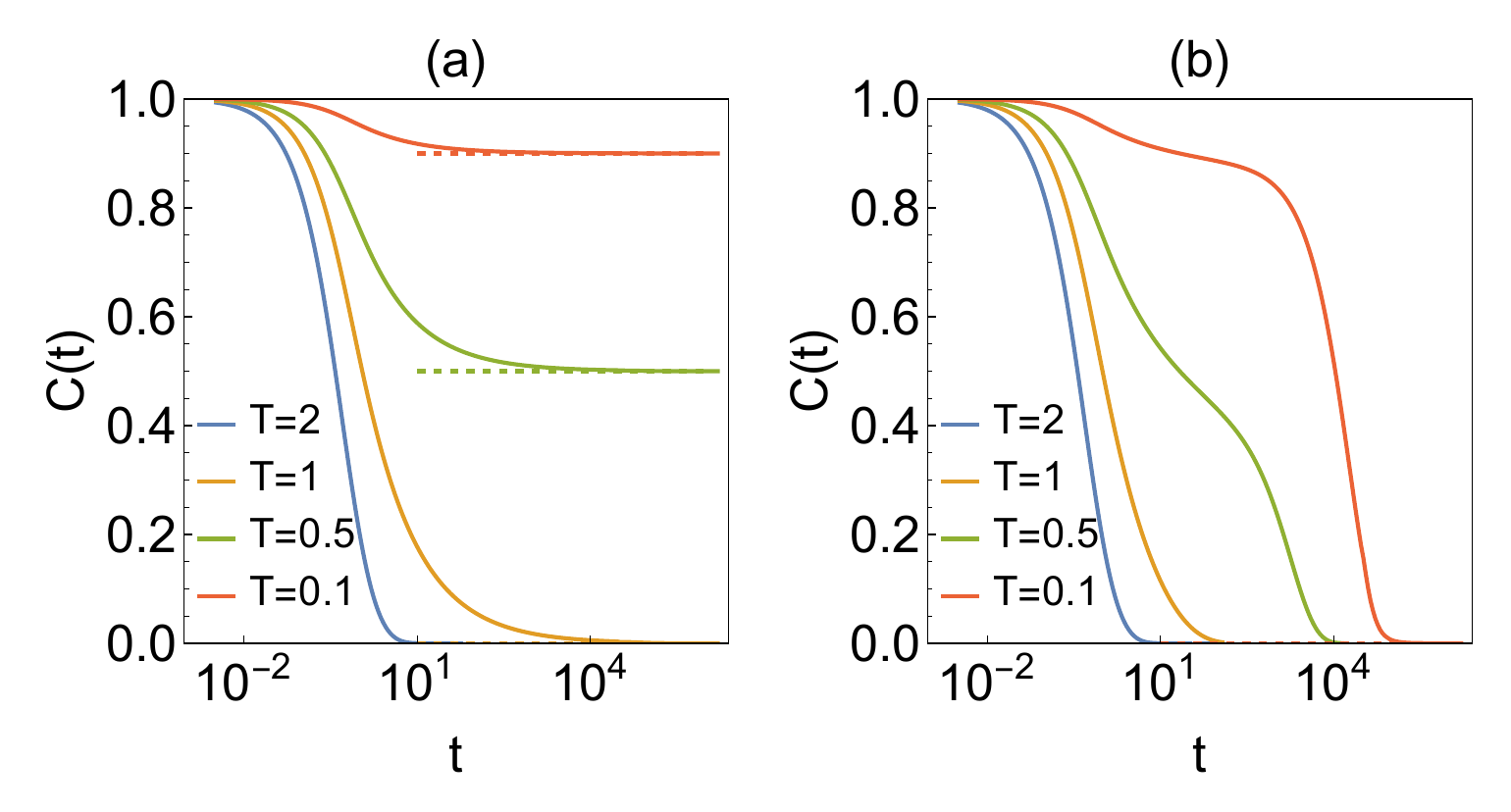} \caption{ Time correlation
 functions in the stationary regime for (a) $\Delta=0$ and (b) $\Delta=0.1$ at several
 temperatures. Here $J=1$. For $\Delta=0$, the correlation approaches a
 nonzero plateau below the transition temperature $T_c=1$ (dashed
 line). For $\Delta>0$, it decays to zero for all temperatures.  }
 \label{154045_5Mar26}
\end{center}
\end{figure}

This analytic conclusion is reflected directly in the long-time behavior
of the correlation function. To illustrate this point,
Fig.~\ref{154045_5Mar26} shows the time correlation function obtained
from
\begin{align}
C(t)=\frac{1}{\pi}\int_0^\infty d\omega\, C(\omega)\cos(\omega t).
\end{align}
For $\Delta=0$, the correlation function approaches a nonzero plateau
below $T_c$ and decays to zero above it. For $\Delta>0$, by contrast,
$C(t)$ decays to zero at long times for all $T$, consistent with the
absence of a finite-temperature glass phase.

Although the transition is removed for $\Delta>0$, the dynamics still
becomes increasingly slow at low temperature. To characterize this slow
relaxation, it is natural to consider the low-frequency limit of the
correlation function,
\begin{align}
\Lambda\equiv \lim_{\omega\to 0} C(\omega).
\end{align}
Since $\Lambda$ is proportional to the time-integrated correlation
function, it provides a natural measure of the characteristic
relaxation time scale. In particular, if the late-time decay is
controlled by a single relaxation time scale,
\begin{align}
C(t)\sim f(t/\tau),
\end{align}
with an integrable scaling function $f$, then one has
$\Lambda\propto\tau$.

\begin{figure}[t]
\begin{center}
 \includegraphics[width=9cm]{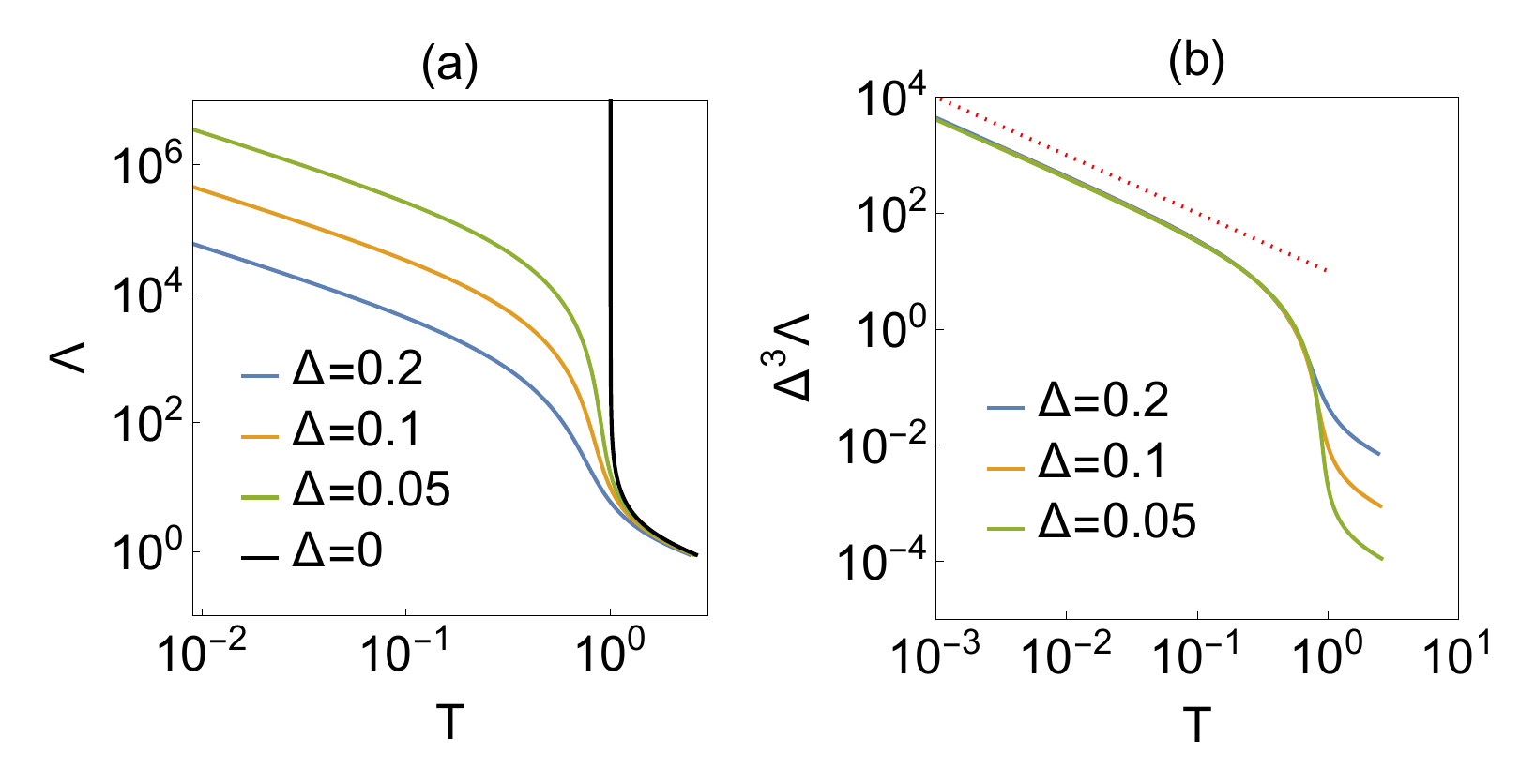}
 \caption{
 (a) $\Lambda=\lim_{\omega\to 0}C(\omega)$ for $J=1$ and several values
 of $\Delta$. For $\Delta=0$, $\Lambda$ diverges at the finite
 transition temperature $T_c=J$, whereas for $\Delta>0$ it diverges only
 as $T\to 0$. (b) Scaling plot of the same data. The red dotted line
 indicates $\Delta^3\Lambda\propto T^{-1}$.
 }
\label{094843_6Mar26}
\end{center}
\end{figure}

\begin{figure}[t]
\begin{center}
 \includegraphics[width=9cm]{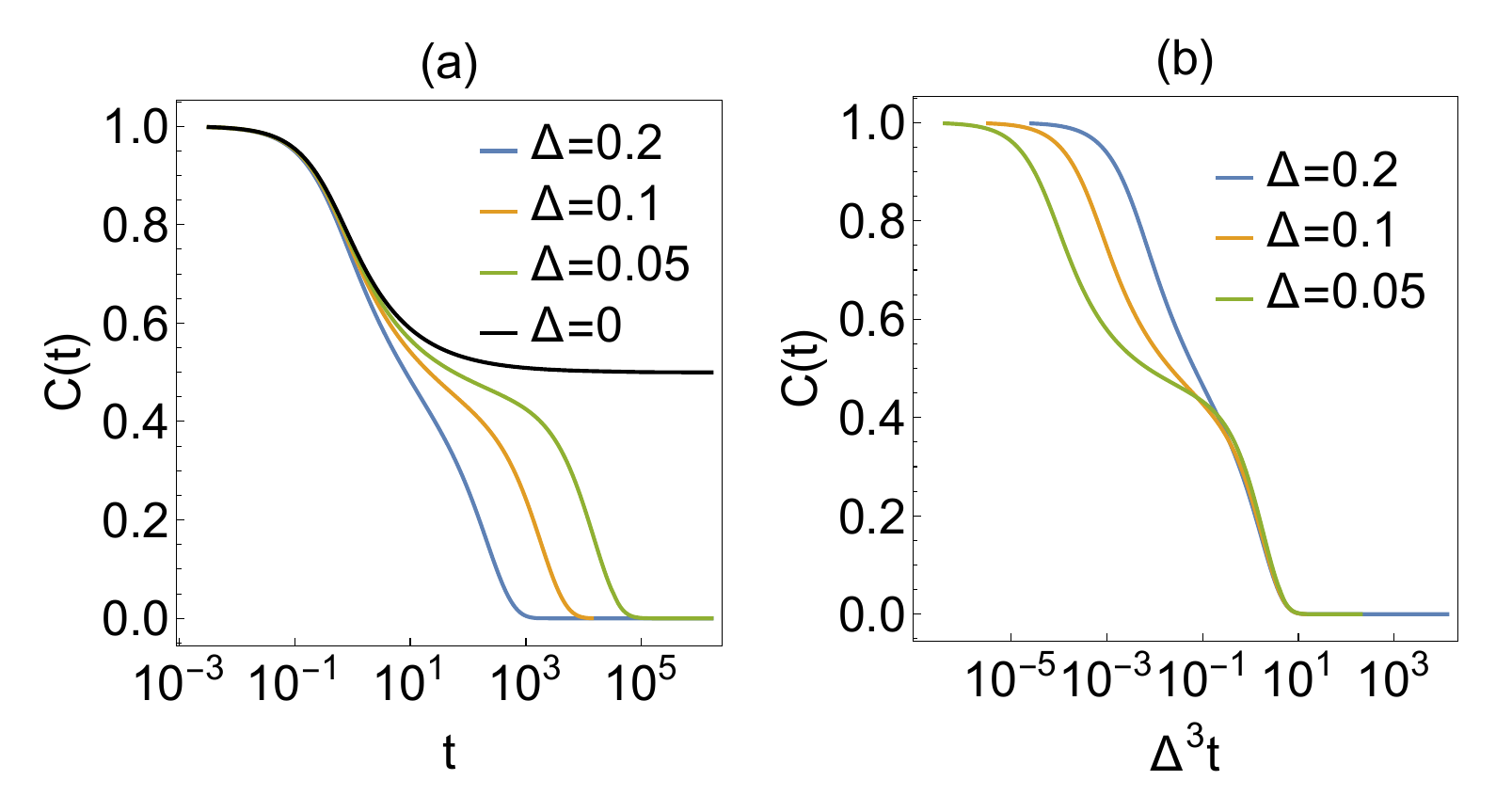} \caption{ (a) Correlation
 functions for several values of $\Delta$ at $J=1$ and $T=0.5$. (b)
 Scaling plot of the same data. The late-time curves collapse when time
 is rescaled by $\Delta^3$.  } \label{171545_27Mar26}
\end{center}
\end{figure}

As shown in Fig.~\ref{094843_6Mar26}~(a), $\Lambda$ diverges at the
finite-temperature transition point for $\Delta=0$, whereas for
$\Delta>0$ it remains finite for all $T>0$ and diverges only in the
limit $T\to 0$. A more detailed analysis, presented in
Appendix~\ref{095601_29Mar26}, shows that
\begin{align}
\Lambda \sim T^{-1}\Delta^{-3}
\end{align}
for small $T$ and small $\Delta$. This scaling is confirmed by the data
collapse in Fig.~\ref{094843_6Mar26}~(b). The above result therefore
shows that the relaxation becomes increasingly slow as $\Delta\to 0$.
For fixed $T>0$, one expects
\begin{align}
\tau\sim \Delta^{-3}
\qquad
(\Delta\to 0).
\end{align}
This expectation is supported by Fig.~\ref{171545_27Mar26}~(a), which
shows that the relaxation becomes progressively slower as $\Delta$
decreases, and more directly by Fig.~\ref{171545_27Mar26}~(b), where the
late-time correlation functions collapse when time is rescaled by
$\Delta^3$.

The divergence of $\Lambda$ as $T\to0$ indicates that the characteristic
relaxation time scale diverges at zero temperature. However, determining
whether the zero-temperature dynamics exhibits aging requires an
analysis of the full two-time dynamics beyond the stationary regime
considered here. We leave this question for future work.

\subsection{General frequency distributions and arbitrary critical frequencies}
\label{general}

We now extend the above result to an arbitrary normalized frequency
distribution $g(\Omega)$, including asymmetric, discrete, and mixed
distributions. We also allow the instability of the stationary solution
to occur at a nonzero frequency $\omega_\ast$.

It is convenient to introduce an auxiliary function 
\begin{align}
Q(\omega)
=
\int d\Omega\,g(\Omega)
\left|R_0(\omega\mid\Omega)\right|^2.
\label{eq:Q_general}
\end{align}
Equation (\ref{215733_2Mar26}) can then be written as
\begin{align}
C(\omega)
=
\frac{2TQ(\omega)}
{1-J^2Q(\omega)}.
\label{eq:C_Q_general}
\end{align}
A stationary solution at $T>0$ requires $J^2Q(\omega)<1$ for all
$\omega$. If the stationary solution becomes unstable at
$\omega=\omega_\ast$, then
\begin{align}
J^2Q(\omega_\ast)
=
J^2
\int d\Omega\,g(\Omega)
\left|R_0(\omega_\ast\mid\Omega)\right|^2
=
1.
\label{eq:instability_general}
\end{align}

A finite-temperature transition requires $Q(\omega)$ to be nonanalytic
at $\omega_\ast$. Indeed, if $Q(\omega)$ were analytic, then near its
maximum it would behave as
\begin{align}
Q(\omega)
=
\frac{1}{J^2}
-
A(\omega-\omega_\ast)^{2}
+\cdots ,\quad 
A>0.
\end{align}
Equation~(\ref{eq:C_Q_general}) would then give
\begin{align}
C(\omega)
\sim
\frac{\mathrm{const.}}
{|\omega-\omega_\ast|^{2}},
\end{align}
which is not integrable and is therefore incompatible with the spherical
constraint at any positive temperature. 

We next ask when $Q(\omega)$ can become nonanalytic. It is useful to
write
\begin{align}
s(\omega)
=
i\omega+\mu-J^2R(\omega),
\end{align}
so that
\begin{align}
R_0(\omega\mid\Omega)
=
\frac{1}{s(\omega)-i\Omega}.
\end{align}
Taking the real part of the response equation gives
\begin{align}
\mathrm{Re}\,R(\omega)
=
\mathrm{Re}\,s(\omega)\,Q(\omega).
\end{align}
Together with
$\mathrm{Re}\,s(\omega)=\mu-J^2\mathrm{Re}\,R(\omega)$, this yields
\begin{align}
\mathrm{Re}\,s(\omega)
=
\frac{\mu}{1+J^2Q(\omega)}>0.
\label{eq:Re_s_positive}
\end{align}
Since $\mathrm{Re}\,s(\omega)>0$, $s(\omega)-i\Omega\neq 0$ for all
$\Omega$, meaning that the $\Omega$ integration in
Eq.~(\ref{eq:Q_general}) introduces no singularity by itself. Thus,
$Q(\omega)$ can become nonanalytic only if $R(\omega)$ becomes
nonanalytic. 
A nonanalyticity of $R(\omega)$ at $\omega_\ast$ requires the
derivatives of the two sides of the self-consistent 
equation~(\ref{110536_2Mar26}) with respect to $R$ to coincide:
\begin{align}
1=J^2\int d\Omega\, g(\Omega)R_0(\omega_\ast\mid\Omega)^2.
\label{eq:response_singularity_general}
\end{align}
Otherwise, the analytic implicit function theorem guarantees that
$R(\omega)$ is locally analytic in a neighborhood of $\omega_\ast$. We
now compare Eqs.~(\ref{eq:instability_general}) and
(\ref{eq:response_singularity_general}). The triangle inequality gives
\begin{align}
\left|
\int d\Omega\,g(\Omega)
R_0(\omega_\ast\mid\Omega)^2
\right|
\leq
\int d\Omega\,g(\Omega)
\left|R_0(\omega_\ast\mid\Omega)\right|^2
=
\frac{1}{J^2}.
\label{eq:triangle_general}
\end{align}
For the response-singularity condition to hold, this inequality must be
saturated. This requires $[R_0(\omega_\ast\mid\Omega)]^2$ to have the
same complex phase for all frequencies in the support of
$g(\Omega)$. Writing $s(\omega_\ast)=a+ib$, we have
\begin{align}
\arg R_0(\omega_\ast\mid\Omega)^2
=2\arctan\left(\frac{\Omega-b}{a}\right),
\end{align}
which varies continuously and strictly with $\Omega$. Therefore,
the equality condition in Eq.~(\ref{eq:triangle_general}) can be
satisfied only when the frequency distribution is supported at a single
point:
\begin{align}
g(\Omega)
=
\delta(\Omega-\Omega_0).
\end{align}
We thus conclude that a finite-temperature glass transition is possible
only in the monodisperse limit. In that case, the common frequency
$\Omega_0$ can be removed by transforming to a frame rotating at
$\Omega_0$, and the model reduces to the conventional spherical
Sherrington--Kirkpatrick model.

For any distribution whose support contains two or more distinct
frequencies, the correlation-instability condition and the
response-singularity condition cannot be satisfied simultaneously.
Consequently, the stationary solution has no finite-temperature glass
transition. This conclusion applies equally to continuous, asymmetric,
discrete, and mixed frequency distributions.

\section{Summary and discussions}
\label{125933_24Mar26}

In this work, we introduced a solvable mean-field model of globally
coupled oscillators with quenched random interactions under a spherical
constraint. For uniform ferromagnetic couplings, the model reproduces
the standard synchronization transition. For fully random symmetric
couplings, we derived closed self-consistent equations for the response
and correlation functions by means of a cavity construction, or
equivalently, a dynamical mean-field theory.

Our main result is that a finite-temperature spin-glass transition
occurs only in the singular monodisperse limit in which all natural
frequencies are identical. In that limit, the model reduces to the
spherical Sherrington-Kirkpatrick model and reproduces its standard
transition. Once the frequency distribution has a finite width or
multiple delta peaks, however, the finite-temperature transition is
suppressed. The physical reason is that frequency dispersion generates a
low-frequency singularity in the correlation function that is
incompatible with the spherical constraint. At zero temperature, by
contrast, the present spherical dynamics still admits a frozen glassy
phase even for finite frequency dispersion.

A concrete prediction that can be tested in randomly interacting
Kuramoto models is the strong slowing down as the monodisperse limit
is approached. In the present model, the characteristic relaxation
time scales as $\tau\sim T^{-1}\Delta^{-3}$. This scaling can be
examined numerically through two-time correlations.

The present results place randomly coupled oscillator models in close
relation to earlier nonequilibrium spherical spin-glass models. In
particular, Crisanti and Sompolinsky showed that perturbations away from
the equilibrium limit suppress the finite-temperature spin-glass phase
while leaving a frozen state at zero
temperature~\cite{crisanti1987}. The present model has a closely related
structure, but the nonequilibrium ingredient is introduced through
distributed natural frequencies rather than through asymmetric
couplings.

The zero-temperature frozen phase should nevertheless be interpreted
with considerable care. In the present spherical dynamics, it survives
even for arbitrarily large frequency dispersion $\Delta$ and arbitrarily
weak coupling $J$. Such behavior is physically implausible for a generic
system of randomly coupled oscillators and strongly suggests that the
residual frozen phase is an artifact of the quasi-linear spherical
approximation rather than a robust property of the original nonlinear
phase dynamics~\cite{crisanti1987}. In a genuinely nonlinear extension,
the oscillator dynamics can feed back on itself and generate additional
self-induced fluctuations even in the absence of external thermal
noise. If these fluctuations carry a continuous low-frequency spectrum,
they may destabilize the frozen state found here. A related lesson has
recently been discussed in nonequilibrium hyperuniform systems, where a
state present in the linear theory is lost once nonlinearities generate
an additional effective large-scale noise
contribution~\cite{Maire2025}. Although the present problem is
different, the same general mechanism may be relevant here:
nonlinearities can qualitatively change the fate of phases that appear
stable within a linearized or spherical description.

It also remains unclear to what extent the spherical approximation can
capture phenomena discussed in previous studies of Kuramoto models with
random interactions, including volcano transitions, chaotic dynamics,
and algebraic decay of the ferromagnetic order
parameter~\cite{daido1992,pruser2024,leon2025dynamics}. Clarifying these
issues remains an important direction for future work.

\begin{acknowledgments}
The author used ChatGPT (OpenAI) to assist with English editing and
improvement of the manuscript text. This project has received JSPS
KAKENHI Grant Numbers 23K13031, and 25H01401.
\end{acknowledgments}

\appendix

\section{Dynamics of the ferromagnetic model}
\label{180637_3Apr26}

Here we investigate the relaxation dynamics of the ferromagnetic model
by directly solving the equation of motion~(\ref{111936_8Jan26}),
starting from the homogeneous initial condition $z_i(0)=1$ for
$i=1,\cdots,N$. In Eq.~(\ref{111936_8Jan26}), the $i$ dependence of
$z_i$ appears only through $\Omega_i$, so that
\begin{align}
z_i(t)=z(t,\Omega_i).
\end{align}
In the thermodynamic limit $N\to\infty$, the order parameter can
therefore be written as
\begin{align}
Z(t)
=
\frac{1}{N}\sum_{i=1}^N z_i(t)
\xrightarrow{N\to\infty}
\int d\Omega\, g(\Omega)\, z(t,\Omega),
\label{211812_8Jan26}
\end{align}
where
\begin{align}
g(\Omega)=\frac{1}{N}\sum_{i=1}^N \delta(\Omega-\Omega_i)
\end{align}
denotes the distribution of natural frequencies.

For the Cauchy distribution, the integral in
Eq.~(\ref{211812_8Jan26}) can be evaluated explicitly by contour
integration in the complex $\Omega$ plane, following the standard
treatment of Lorentzian frequency distributions in coupled-oscillator
models~\cite{ott2008}. If $z(t,\Omega)$, viewed as a function of
complex $\Omega$, is analytic in the upper half-plane and sufficiently
well behaved at infinity, the contribution from the large semicircle
vanishes. Since $g(\Omega)$ has a single pole at $\Omega=i\Delta$ in
the upper half-plane, the residue theorem yields
\begin{equation}
Z(t)=\int d\Omega\, g(\Omega)\, z(t,\Omega)=z(t,i\Delta).
\label{A5}
\end{equation}
Evaluating the equation of motion at $\Omega=i\Delta$, we obtain
\begin{equation}
\dot Z(t)=\left(-\mu-\Delta+K\right)Z(t).
\end{equation}
Using the spherical constraint, $\mu(t)=K|Z(t)|^2+T$, we finally arrive
at
\begin{equation}
\dot Z(t)=(K-T-\Delta)Z(t)-K|Z(t)|^2Z(t).
\end{equation}
In the steady state, $\dot{Z}=0$, and therefore
\begin{align}
\abs{Z}=
\begin{cases}
0, & T>K-\Delta,\\[3pt]
\sqrt{\frac{K-T-\Delta}{K}}, & T\le K-\Delta,
\end{cases}
\end{align}
which is consistent with the result presented in the main text.

\section{Asymptotic of $\Lambda$}
\label{095601_29Mar26}
Here we discuss the scaling behavior of $\Lambda=\lim_{\omega\to 0}C(\omega)$
for small $\Delta$ and $T<J$.
For this purpose, we expand the numerator and denominator
of Eq.~(\ref{112733_28Mar26})
for small $\omega$ and $\Delta$:
\begin{align}
 C(\omega)= 
 \frac{2T\alpha(\omega)}{\delta\mu + 2J^2(\alpha(0)-\alpha(\omega))}
\approx \frac{2T\alpha(0)}{\delta\mu + \omega^2\Delta^{-3}},\label{112847_28Mar26}
\end{align}
where
\begin{align}
\delta \mu = \mu-\mu_c.
\end{align}
The spherical constraint is now
\begin{align}
 1 = 2T\alpha (0)\int d\omega \frac{1}{\delta \mu + \omega^2 \Delta^{-3}}
 \propto T \sqrt{\frac{\Delta^3}{\delta\mu}},
\end{align}
leading to 
\begin{align}
\delta\mu \propto  T^2 \Delta^3.
\end{align}
Substituting it back into Eq.~(\ref{112847_28Mar26}), we get 
\begin{align}
 \Lambda \propto T^{-1}\Delta^{-3}.
\end{align}
This scaling implies that if $\Delta^3 \Lambda$ is plotted as a function
of $T$, the plots for different $\Delta$ collapse on a single master
curve, which is inversely proportional to $T$. Scaling plot in
Fig.~\ref{094843_6Mar26}~(b) indeed verifies this conjecture.



\bibliography{reference}

\end{document}